\documentclass[twoside]{article}
\usepackage{amsmath, amsfonts, amssymb}
\usepackage{graphics}
\usepackage{Aat}
\usepackage{tabularx}
\usepackage{Aattable}

\newcommand{\FigBox}[2][\columnwidth]{\framebox[#1]{\rule{0pt}{#2}}}

\begin{document}

\thispagestyle{myfirst}

\setcounter{page}{1}

\mylabel{00}{4} \mytitle{Gaseous Disks of Spiral Galaxies: Arms
and Rings} \myauthor{Edward Liverts$^{1,2}$, Evgeny Griv$^{2}$,
Michael Gedalin$^{2}$, David Eichler$^{2}$ and Chi Yuan$^{3}$}
\myadress{$^{1}$ Department of Theoretical Physics, Novgorod State
University, Novgorod 173003, Russia,\\ $^2$ Dept. of Physics,
Ben-Gurion University, Beer-Sheva 84105, Israel\\ $^3$ Academia
Sinica Institute of Astronomy, Taipei 11529, Taiwan\\ E-mail
eliverts@bgumail.bgu.ac.il} \mydate{(Received December  1, 2000)}
\myabstract{An improved linear stability theory of
small-amplitude oscillations of a self-gravitating,
infinitesimally thin gaseous disk of spiral galaxies has been
developed.  It was shown that in the differentially rotating
disks for nonaxisymmetric perturbations Toomre's modified
critical $Q$-parameter is larger than the standard one.  We use
hydrodynamical simulations to test the validity of the modified
local criterion.} \mykey{Galaxies: kinematics and dynamics,
structure--instabilities}

\section{Results}

Lin {\it et al.} (1969) and Shu (1970) developed
a linear theory of tightly wound density waves to solve the
problem of the spiral structure in disk galaxies.
An important discriminant in the original
Lin--Shu dispersion relation which connects frequency and wavenumber
of excited waves is Toomre's local stability parameter $Q$ (Toomre,
1964): when $Q \ge 1$,
the self-gravitating disk is stable against axisymmetric
(ringlike) perturbations.  This local criterion gives a
necessary condition for radial stability.  It does not obviously
address the stability of nonaxisymmetric modes.
Lau and Bertin (1978), Lin and Lau (1979),
Morozov (1980), Polyachenko (1989)
and Griv {\it et al.} (1999) extended
the original works of Lin {\it et al.}
(1969) and Shu (1970) by including the azimuthal forces.  It was
demonstrated that the presence of the differential rotation (or
shear) results in quite different dynamical properties
of the axisymmetric and nonaxisymmetric perturbations.
A dispersion relation for arbitrary perturbations has
been rederived.  This generalized Lin--Shu--Kalnajs dispersion
relation leads to the following modified local stability criterion
obtained, e.g. by Morozov (1980) and Griv {\it et al.} (1999):
\begin{equation}
Q \ge \{ 1 + [ ( 2 \Omega / \kappa
)^2 - 1 ] \sin^2 \psi \}^{1/2} ,
\end{equation}
\noindent
where the condition $2 \Omega / \kappa > 1$ always holds in the
differentially rotating system.  The quantities $\Omega(r)$ and
$\kappa (r)$ denote the angular velocity of galactic rotation
and the epicyclic frequency, respectively, at the distance $r$
from the center.  The pitch angle $\psi$
between the direction of the wave front and the tangent to the
circular orbit in Eq. (1) is $\psi = \mbox{arctan} (m / k_r r)$,
where $m$ is the
number of spiral arms, and $k_r$ and $k_\varphi = m/r$
are the radial and the azimuthal wavenumbers.
The parameter $\{1 + [(2 \Omega/\kappa)^2 - 1]
\sin^2 \psi \}^{1/2}$ is an additional stability
parameter which depends on both the pitch angle and the amount of
differential rotation in the galaxy.

\begin{figure}[t]
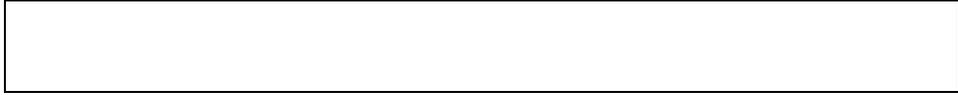

\FigBox{1cm}
\caption{The
nearly-isothermal differentially rotating system with $Q=1$ is
found to be unstable with respect to almost circular
perturbations growing on a dynamical time scale. Eventually, the
almost circular structure converts into a one-armed spiral.  The
underlying potential in a large fraction of spiral galaxies, e.g.
in M101, is now believed to have this lopsided form.}
 \end{figure}

It is clear from the modified criterion (1) that in
a nonuniformly rotating disk
for nonaxisymmetric perturbations the
modified stability criterion is larger than $Q=1$
(although still of the order of unity).  A relationship
exists between Eq. (1) and what Toomre (Binney
and Tremaine, 1987) called ``swing amplification."
The free kinetic energy associated
with the differential rotation of the system under study is one
possible source for the growth of the energy of these spiral
perturbations, and appears to be released when
angular momentum is transferred outward.

As one can see from Eq. (1), the critical stability parameter
grows with $\psi$.  Consequently, in order to suppress the most
``dangerous," in the sense of the loss of gravitational stability,
open perturbations ($\psi \stackrel{>}{\sim} 45^\circ$),
$Q$ should obey the following modified criterion:
\begin{equation}
Q \ge Q_{\mathrm{mod}} = 2 \Omega / \kappa .
\end{equation}
\noindent
One should keep in mind that equation (2) is clearly only an
approximate one.  It is clear, however,
that the modified criterion for the
local stability of a gaseous disk against arbitrary Jeans
perturbations should be approximately of the form of Eq. (2):
the condition for the spiral Jeans instability onset is $Q <
2.0-2.5$ at all radii.

The value of Toomre's stability parameter $Q$ is critically
important for any gravitational theory of spiral structure in
galaxies.
The modified local stability parameter has been discussed at
length by Griv and Peter (1996) and
Griv {\it et al.} (1999), using a kinetic approach.
Here we discuss the problem using a gasdynamic approximation.

\begin{figure}[t]
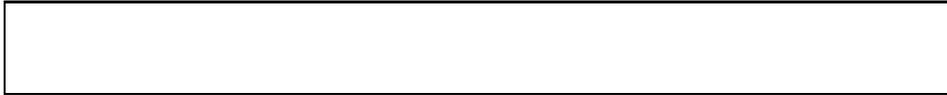

\FigBox{1cm}
\caption{Time-development of
Toomre's $n=4$ gaseous model with $Q=2$.  In agreement with the
theory, all Jeans type perturbations are suppressed including the
most unstable nonaxisymmetric ones.}
 \end{figure}

The main objectives of the current work is to check the modified
local stability parameter (2) numerically using the method of
hydrodynamical simulations.
In particular, we focus on the thermal
motion effect in differentially rotating disks.
A nonuniformly rotating gaseous disk is considered.
Angular velocity of the disk is defined by the stellar gravitational
potential because the mass of gas is much smaller the stellar system.
We solve the gasdynamical equations numerically in two-dimensions by
using an Eulerian hydro-code.  The hydrodynamic part of the basic
equations is solved by the implicit ``upstream--downstream" scheme.
At the symmetry axis we require that no mass flows through the axis.
At the other boundaries we require that the gas should not be able
to leave the grid.  We use
$1024 \times 1024$ Cartesian grid points covering a $20 \times 20$
kpc region.  Therefore, the spatial resolution is about 20 pc.
A periodic Green function is used to calculate the self-gravity.
The initial conditions are an axisymmetric and rotationally supported
disks.  Small random density and velocity fluctuations
are added to the initial system.
The problem consists in calculating the reaction of the gas to the
fluctuations.  (A time $t=1$ is taken to correspond to a single
revolution of the initial disk.)

In Figure 1 we show a series of
snapshots from a run with the cool model,
namely, Toomre's $n=4$ model (Binney and Tremaine, 1987, p. 44),
in which $Q$-value is equal to 1 over the whole system.
As has been predicted in the theory,
the spiral Jeans instability develops quickly in the system
during the time of the first rotation.
A typical wavelength (a typical distance between the spirals) is
$\sim 2 \lambda_J$, indicating that perturbations
with Jeans--Toomre wavelength $\sim \lambda_J$ have the fastest growth
rate.  Such a size of a density wave is in agreement with
the theory (Griv {\it et al.}, 1999).

In the second set of experiments with Toomre's $n=4$ disk,
we set at $t=0$ the parameter $Q=2 \Omega / \kappa \approx 2$.
The results of the experiment are shown in
Figure 2.  As one can see, the system becomes
hydrodynamically stable.  The result agrees
with the theoretical explanation described in the paper.

We conclude that in agreement with the theory, the disks
become progressively more stable as Toomre's
$Q$-parameter grows.  In order to suppress the
instability of arbitrary Jeans type
perturbations in a differentially rotating gaseous disk,
Toomre's $Q$-value must exceed $Q_{\mathrm{mod}}$ (Eq. [2]).

\subsection*{References}

\rf{Binney J. and Tremaine S.: (1987), {\it Galactic Dynamics},
   Princeton Univ. Press.}

\rf{Griv E. and Peter W.: (1996), {\it Astrophys. J.}
   {\bf 469}, 89.}

\rf{Griv E., Rosenstein B., Gedalin M., and Eichler D.: (1999),
   {\it Astron. Astrophys.} {\bf 347}, 821.}

\rf{Lau Y.Y. and Bertin G.: (1978), {\it Astrophys. J.} {\bf 226}, 508.}

\rf{Lin C.C., Yuan C., and Shu F.H.: (1969), {\it Astrophys. J.}
   {\bf 155}, 721.}

\rf{Lin C.C. and Lau Y.Y.: (1979), {\it Stud. Appl. Math.}
   {\bf 60}, 97.}

\rf{Morozov A.G.: (1980), {\it Soviet Astron.} {\bf 24}, 391.}

\rf{Polyachenko V.L.: (1989), in {\it Dynamics of Astrophysical
   Discs}, ed. Sellwood J.A., Cambridge Univ. Press, p. 199.}

\rf{Shu F.H.: (1970), {\it Astrophys. J.} {\bf 160}, 99.}

\rf{Toomre A.: (1964), {\it Astrophys. J.} {\bf 139}, 1217.}

\end{document}